# Slip viscosity and strain-rate viscosity in Taylor-Couette laminar flows: Experimental falsification and end-wall effects

Jian He(何建) [1], Jin Wang(王津) [1], Qiaocong Kong(孔巧聪) [1], Penglong Zhao(赵朋龙) [1], Xiaoshu Cai(蔡小舒) [2], Xiaohang Zhang(张晓航) [3], Wennan Zou(邹文楠) [1]✉

[1] Institute of Fluid Mechanics/Institute for Advanced Study/College of Advanced Manufacturing, Nanchang University, Nanchang 330031, China;

[2] School of Energy and Power Engineering, University of Shanghai for Science and Technology, Shanghai 200093, China;

[3] School of Science, Jiangxi University of Water Resources and Electric Power, Nanchang, 330099, China.

✉Wennan Zou, **Email:** zouwn@ncu.edu.cn

**ABSTRACT.** The viscous force should be shear force, the difference between the strain-rate viscosity and the slip viscosity is that the former has conjugate shear force, while the latter does not. The study in this paper verifies the physical authenticity of two viscosity models through Taylor Couette laminar flow experiments with inner and outer cylinders rotating at the same angular velocity, and numerically investigate the influence of relative cylinder spacing and rotational speed on the circumferential velocity under the slip model. The experimental results of LDV measurement with a relative cylinder spacing of 0.3 indicate that the maximum deviation from rigid-body rotation is about 0.86%, which is consistent with the theoretical prediction of slip viscosity model. The numerical simulations show that the end-walls have no effect under the strain-rate viscosity model; but when the slip viscosity model is introduced, the end-walls inevitably bring about the circumferential velocity profile changing along the axial direction, and result in a three-dimensional (3D) spiral streamline pattern influenced by the relative cylinder spacing and angular speed of cylinders.

## 1. INTRODUCTION

Newton (1687)[1] proposed that "the resistance arising from the want of lubricity in the parts of a fluid, is, other things being equal, proportional to the velocity with which the parts of the fluid are separated from each other", which forms the conceptual foundation of slip viscosity. but Newton gave no mathematical formula for his viscous friction. Euler (1755)[2] built up an excellent flow theory for the fluid that can sustain no shearing force. D'Alembert's paradox of 1768 (cf. Darrigol, 2005[3]) pointed out the fact that the viscosity must be taken into account for real fluids. Through the unremitting efforts (Darrigol, 2002[4], 2005) of Navier in 1822, Cauchy in 1823, Poisson in 1829 and Saint-Venant in 1937, Stokes in 1845 [5] finally established the strain-rate viscosity model by analogy with the elastic stress. For air and water, it is surprising that the linear law is found to be accurate over a very large range of values of speed difference which includes the values commonly met in practice (Batchelor, 2000 [6]).

For almost 180 years now, almost no one would have thought that there could be a different model for the viscous friction inside fluids, although the microscopic mechanisms of gas viscosity and liquid viscosity have been clearly elucidated (Kreuzer, 1981 [7]). As for the Couette flow of incompressible fluids between two parallel plates (Fig. 1, left), the strain–rate viscosity model formulates the total viscous stress in the form (Fig. 1, left, bottom)

$$\boldsymbol{\sigma} = \mu\frac{U}{h}(\mathbf{e}_1 da_2 + \mathbf{e}_2 da_1), \tag{1}$$

while the viscous friction given by the slip viscosity model is

$$\boldsymbol{\sigma} = \mu \frac{U}{h} \mathbf{e}_1 da_2. \tag{2}$$

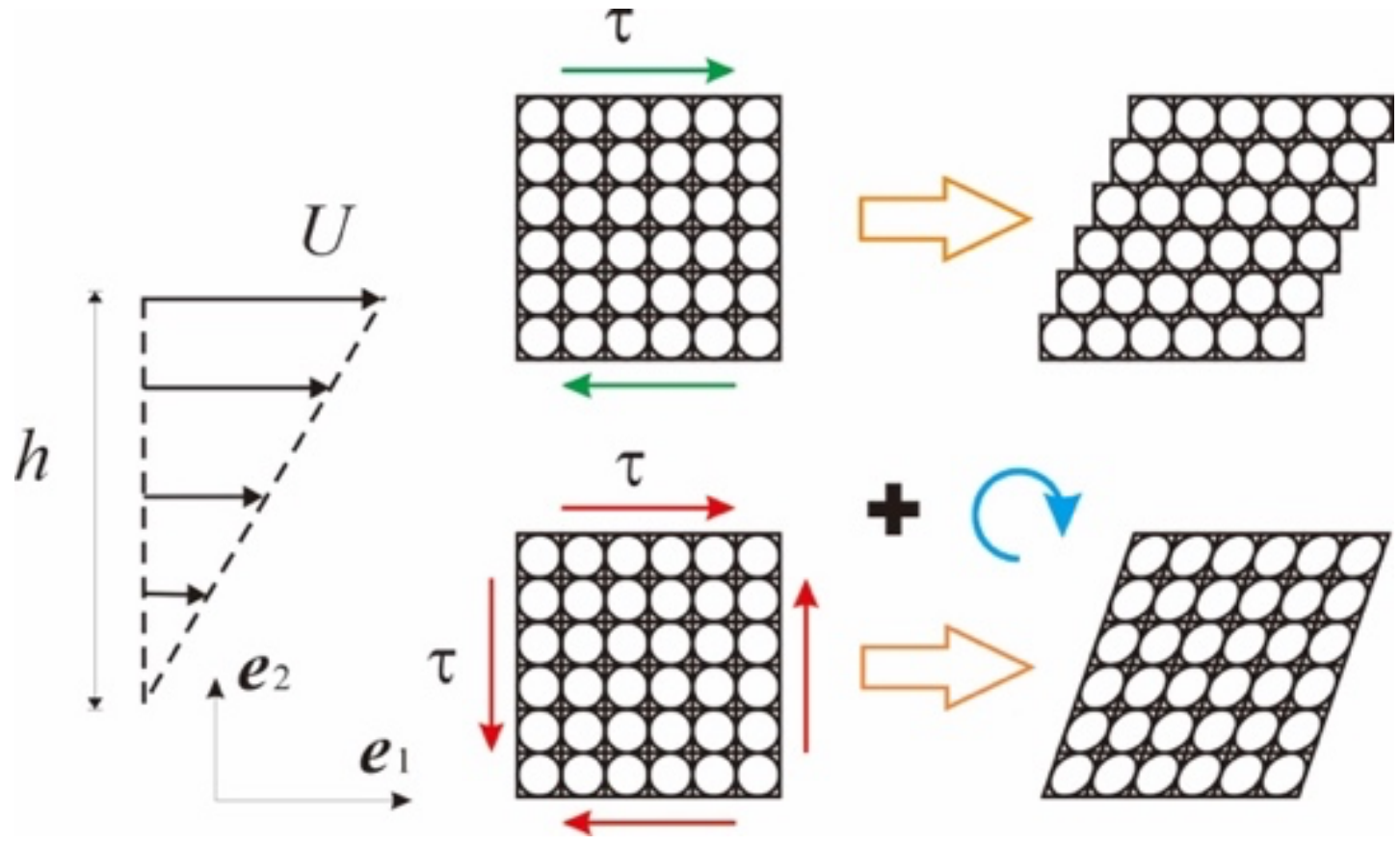

**Figure 1.** Slip (top) and deformation (bottom) ideas of the Couette flow between two parallel plates

The generalization of (1) using the strain–rate is well-known to be

$$\sigma_{ji} = 2\mu S_{ij}, S_{ij} = \tfrac{1}{2}\left(\partial_j v_i + \partial_i v_j\right). \tag{3}$$

But for the slip viscosity model, it has to nontrivially introduce the spatial connection (Chern *et al*., 2000 [8]; Edelen, 1985 [9]) or the swirl field $A_k^i$ (Zou, 2003, 2017)[10,11] to remove the curvature effect of streamlines, say

$$\sigma_{ji} = \mu D_j v_i = \mu\left(\partial_j v_i + \epsilon_{ilm} A_j^l v_m\right), \tag{4}$$

where $D_j$ indicates the covariant derivative replacing the ordinary derivative $\partial_j$ in the consideration of viscous interactions, $\epsilon_{ilm}$ is the permutation symbol. For laminar flows, let $v_i = V n_i, V = \sqrt{v_i v_i}$, the swirl field can be described by the direction field $n_i$ as

$$A_j^i = -\epsilon_{ipq} n_p \partial_j n_q \tag{5}$$

which results in

$$\sigma_{ji} = \mu D_j v_i = \mu n_i \partial_j V. \tag{6}$$

Continuing to derive the integration of viscous surface forces using the covariant derivative defined by the swirl field, the laminar flow equation under the slip viscosity model can be derived to be [11]

$$\frac{dv_i}{dt} = \frac{\partial v_i}{\partial t} + v_j \partial_j v_i = -\partial_i \frac{p}{\rho} + \nu n_i \nabla^2 V, \tag{7}$$

which is different from the famous Navier–Stokes (N–S) equations [6].

$$\frac{dv_i}{dt} = \frac{\partial v_i}{\partial t} + v_j \partial_j v_i = -\partial_i \frac{p}{\rho} + \nu \nabla^2 v_i \tag{8}$$

under the strain–rate viscosity model.

Consider the Taylor-Couette (T-C) flow, namely the flow of fluid between two concentric cylinders that are standing upright and infinitely long, with radii $R_1$ and $R_2$, and rotating around their axis with angular speed $\Omega_1$ and $\Omega_2$, respectively (Taylor, 1935, 1936 [12]; van Gils et al., 2011 [13]; Huisman et al., 2012 [14]). The fluid is assumed to be incompressible and have linear viscosity; its flow is laminar, not considering gravity or incorporating gravity into pressure. Then, under the assumption that both the axis of cylinders and the direction of gravity are in the $z$-axis of the cylindrical coordinate system ($\boldsymbol{r}$, $\theta$, $z$), the flow fields with properties

$$v_r = v_z = 0, v_\theta = v_\theta(r), p = p(r) \tag{9}$$

must satisfy the boundary conditions

$$v_\theta = \Omega_1 R_1, r = R_1; v_\theta = \Omega_2 R_2, r = R_2. \tag{10}$$

The analytic solutions from (7) and (8)

$$V_{\text{Slip}} = \Omega_1 R_1 + (\Omega_2 R_2 - \Omega_1 R_1)\frac{\ln r - \ln R_1}{\ln R_2 - \ln R_1}, V_{\text{Strain-rate}} = \frac{\Omega_2 R_2^2 - \Omega_1 R_1^2}{R_2^2 - R_1^2} r + \frac{\Omega_2 - \Omega_1}{R_2^{-2} - R_1^{-2}} r^{-1}, \quad (11)$$

respectively. Back to Newton's thought experiment without the outer cylinder, it is obvious that the solution $V_{\text{Slip}} = \Omega_1 R_1$ instead of $V_{\text{Strain-rate}} = \Omega_1 R_1^2 r^{-1}$ supports Newton's conclusion (Corollary I in Section IX, Book II of the Principia). Smith (1998)[15] argued that "Newton mistakenly balanced the forces on the inside and outside of each thin fluid shell surrounding the body, not the torques". Rather say, Newton recognized the slip viscosity model rather than the strain–rate viscosity model.

In this paper, we will testify which viscosity model is false through the experiment of the T-C laminar flow with the two cylinders rotating at the same constant angular speed $\Omega$. When $\Omega_1 = \Omega_2 = \Omega$, the velocity formulae in (11) are simplified to be

$$V_{\text{Slip}} = \Omega R_1 + \Omega(R_2 - R_1)\frac{\ln r - \ln R_1}{\ln R_2 - \ln R_1}, V_{\text{Strain-rate}} = \Omega r. \quad (12)$$

Introducing the orbital angular speed by $\omega = v_\theta(r)/r$, the formula

$$\frac{\omega}{\Omega} = \frac{V_{\text{Slip}}}{V_{\text{Strain-rate}}} = \frac{\ln(1+\eta) + \eta\ln(1+\delta\eta)}{(1+\delta\eta)\ln(1+\eta)} \quad (13)$$

can be used for the verification of experimental results, where $\eta \equiv (R_2 - R_1)/R_1$ is the relative cylinder spacing (RCS), and $\delta \equiv (r - R_1)/(R_2 - R_1)$ the dimensionless radial distance (DRD). The theoretical curves for different RCSs $\eta$ to be testified through the experimental results are shown in Fig. 2.

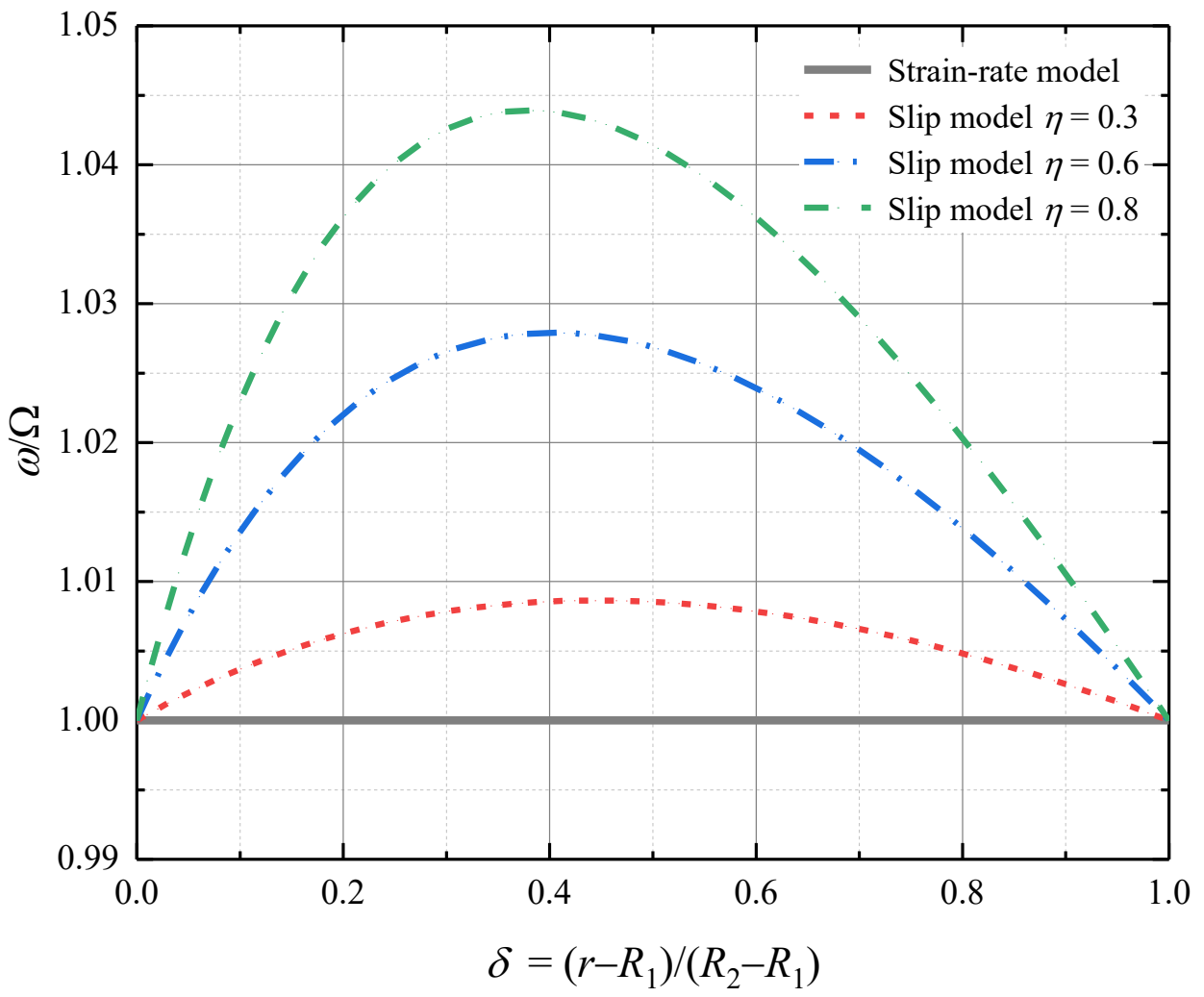

**Figure 2.** The normalized orbital angular speeds $\omega/\Omega$ under different relative cylinder spacing (RCSs) $\eta \equiv (R_2 - R_1)/R_1$ are functions of dimensionless radial distance (DRD) $\delta \equiv (r - R_1)/(R_2 - R_1)$.

The theoretical predictions in (13) assume that the lengths of cylinders are infinite; in practice, the end-walls are inevitable. The existence of end-walls would have no effect to the prediction of the strain–rate viscosity model since the boundary conditions of end-walls coincide with the rigid-body rotation solution. But the circumferential velocity profile as the rigid-body rotation conflicts with the theoretical prediction of the slip viscosity model without end-walls, implying that a change of velocity profile along the axial direction must happen. We will make use of numerical simulation of laminar T-C flows to investigate the effects of end-walls.

The remainder of this paper is arranged as follows. In Section 2, we report the measurement results from a set of T-C laminar flow experiments with RCS $\eta = 0.3$ and the velocity profile measured by LDA (Durst, 1981 [16]; Albrecht *et al*., 2003 [17]; Zhang, 2010 [18]). In Section 3, the simulations using the ANSYS Fluent software are reported, where use is made of the UDF and the three-dimensional helical characteristics of flows due to the end-walls are analyzed. The comparison between two viscosity models and the experimental results are presented with analyses. Finally in Section 4, a brief summary and outlook is given.

## 2. EXPERIMENTAL RESULTS

Our study aims to distinguish two viscosity models, so that we need to conduct a curved laminar flow where the curvature of streamlines significantly affects the distribution of velocity profiles. The simplest curved flow is no doubt the T-C flow between two concentric cylinders where all streamlines are circle according to the N-S equations. In practice, the cylinders have to be length finite. In order to avoid the discontinuity in boundary velocity, we adopt the plan in which the cylinders and the end-walls rotate at the same angular velocity. Once the experimental platform of the T-C flow is built up, the next is to precisely measure the velocity profile when the flow is in a stable laminar flow state. We will use LDA to measure the velocity profile on the axial center section.

### 2.1. Experimental setup

Fig. 3 shows the schematic diagram and the photograph of experimental set-up, including the experimental platform and the measurement system (Wang, 2020 [19]; He, 2024 [20]).

The experimental platform consists of a stainless-steel base, a direct drive motor, inner and outer aluminum alloy cylinders, and an upper cover. The base, motor, and cylinder are connected by high-strength bolts and embedded methods through three flanges (two aluminum alloy and one stainless-steel). The upper cover and cylinders are compressed with a threaded block through the threaded thin column extending from the inner cylinder, and sealed with rubber pads and glass glue. The direct drive motor with model M-PS3060KN002 is provided by the NSK Company. The end cover made of K9 glass is a circular disk with 140 mm diameter, 10 mm thickness, and a depth of 3 mm embedded in the outer cylinder. The solid inner cylinder has a height of 490 mm and a diameter of $D_1 = 100$ mm, while the hollow outer cylinder has an inner diameter of $D_2 = 130$ mm, a wall thickness $t = 5$ mm, a height of 498 mm, and a depth of 5 mm embedded in the flange. During the assembly process of the experimental platform, it is necessary to ensure that all surfaces of the flanges are horizontal, the two cylinders have good coaxiality, and their axis of symmetry is perpendicular. Some geometric parameters are indicated in Fig.3($a$).

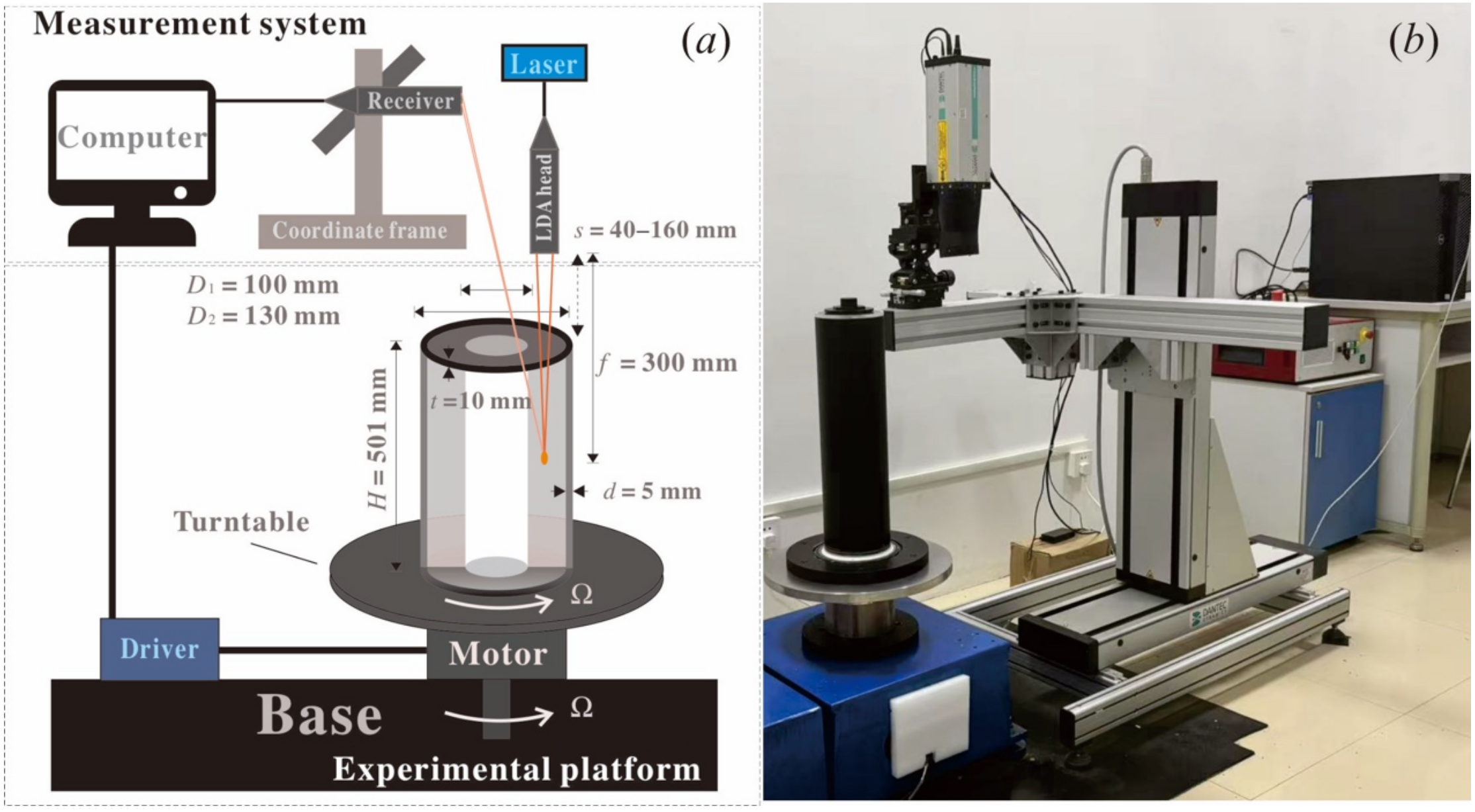

**Figure 3**. ($a$) Schematic diagram and ($b$) physical picture of the experimental platform

The distilled water, in which the tracer particles are added, filled in the gap between cylinders. The tracer particles, about 5 mg in a liter of water and mixed well, are hollow glass beads with a density of 1.03 $g/cm^3$, a refractive index of 1.5 and an average nominal diameter of 10 μm, which can offer good scattering efficiency and a sufficiently small velocity lag. The measurement system in this experiment is the LDA equipped with a two-dimensional (2D) coordinate frame, produced by the Dantec Company. The specific parameters of the LDA system are shown in Table 1. The laser head is installed on a coordinate frame capable of achieving high-precision position. The axis of symmetry of two cylinders should be parallel to the coordinate plane of frame, while the plane formed by the two visible lasers emitted by the laser head should be perpendicular to the coordinate plane of frame.

Table 1. LDA parameters used in present measurement

| LDA | Symbol | Value |
|---|---|---|
| Wavelength [nm] | $\lambda$ | 660 (Red) / 785 |
| Beam diameter [mm] | $d_l$ | 2.6 |
| Beam spacing [mm] | $2w$ | 60 |
| Probe focal distance [mm] | $f$ | 300 |
| Number of fringes | $N$ | 29 |
| Fringe spacing [μm] | $\Delta x_{\mathrm{a}}$ | 3.287 |
| Beam half-angle [°] | $\alpha_{\mathrm{a}} = 0.5\theta_{\mathrm{a}} \approx \dfrac{180w}{\pi f}$ | 5.711 |
| Probe volume dX [mm] | $d_x = d_f/\cos\alpha_{\mathrm{a}}$ | 0.09745 |
| Probe volume dY [mm] | $d_y = d_f$ | 0.09696 |
| Probe volume dZ [mm] | $d_z = d_f/\sin\alpha_{\mathrm{a}}$ | 0.9745 |

There are two pairs of laser beams with two different wavelengths $\lambda = 660$ (Red) / 785 nm emitted by the laser head. We choose the first-dimensional laser beams to measure the circumferential velocity profile at the different axial sections. The ellipsoidal probe volume formed by the two cross beams has three semi-axes $\{\rho_r, \rho_\theta. \rho_z\} = \{d_x/2, d_y/2, d_z/2\}$. Since water and glass have different refractive indices of light compared to air. The intersection point of the two laser beams will move from A to B, as shown in Fig. 4. As the distance from the laser head to the top surface of the end cover is set to be $s$, the coordinate $z_{\mathrm{B}}$ of measurement point is

$$z_B = H - t - \left(\frac{f-s}{n_{\mathrm{a}}} - \frac{t}{n_{\mathrm{g}}}\right) n_{\mathrm{w}}, \tag{14}$$

where

$$n_{\mathrm{a}} = 1, n_{\mathrm{w}} = 1.333, n_{\mathrm{g}} = 1.514, t = 10\ \mathrm{mm}, f = 300\ \mathrm{mm}, H = 501\ \mathrm{mm}. \tag{15}$$

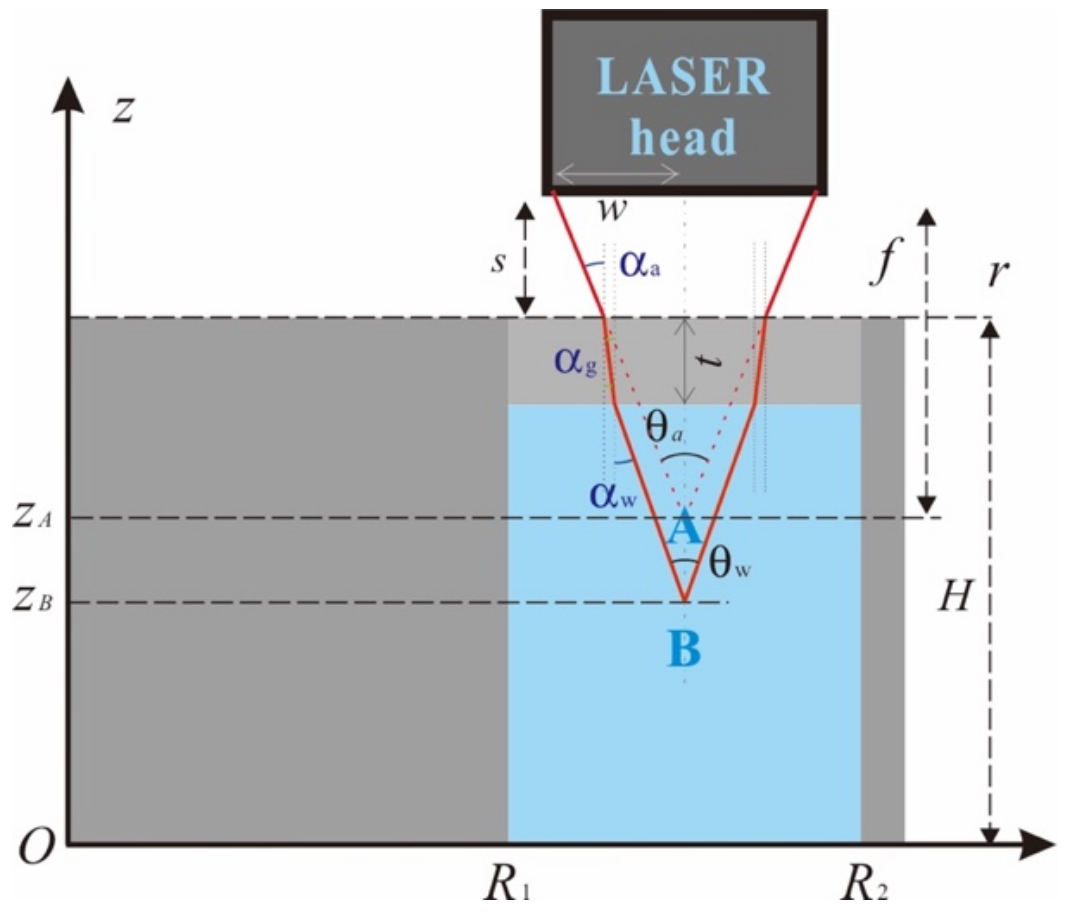

**Figure 4.** Optical paths for the velocity measurement in this experiment

The calibration point for the radial position of the probe volume is the axis of symmetry of the cylinders, while the axial position of the probe volume is calibrated through the upper surface of the flange connected with the cylinders. Due to the interference from the protruding part of the inner cylinder, the distance $s$ has to be large than 40 mm. From the formula (14), that means the measurable axial range of the water body is $z_B = 154 - 490$ mm. Along the radial direction, due to the influence of the cylinder walls and the width of two laser beams, the measurable and effective range becomes $r = 52 - 60$ mm.

### 2.2. Measurement signals and results

During the experiment, the laboratory temperature is to 25 ℃ with a resolution of 1 ℃, and the laboratory humidity is to 60%, all controlled by the dehumidification mode of air conditioner. In order to form a steady laminar flow, the experimental platform needs to be continuously powered on and rotate stably at a lower Reynolds number for about 12 hours to begin the measurement. In the present experiment, the range of angular velocity is set to be $0.13 - 0.15$ Hz, while the uncontacted LDV system can be used to monitor the state of the flow.

Before the measurement, place the laser head above the end cover of the concentric cylinders and adjust the symmetrical vertical plane of the two visible lasers to include the axis of symmetry of the cylinders. At each measurement point, the signals of circumferential velocity are recorded by an LDV laser at a frequency of 8 Hz. After the process by the Dantec Dynamics system, the velocity resolution can be calibrated to be below 0.01 mm/s. Under the condition that the efficient analysis exceeds 99.5%, we adopt 2000 sampling signals for each measurement point.

The uncertainty of measurement signals may stem from multiple factors, including the geometric deviation of the experimental setup and the optical deviation of the measuring device. In order to verify the stability and repeatability of measurement, we compare the measurement signals at least 3 times under the same conditions for each measurement point. After screened with a 95% confidence level, Fig. 5 presents the original signal segment of the measurement point signal at radial position $r = 55$mm and the axial center section $z_B = 247$ mm of the water body, and its analysis diagram. In the flow of Fig. 5, the angular speed is taken to be $\Omega = 0.15$ Hz.

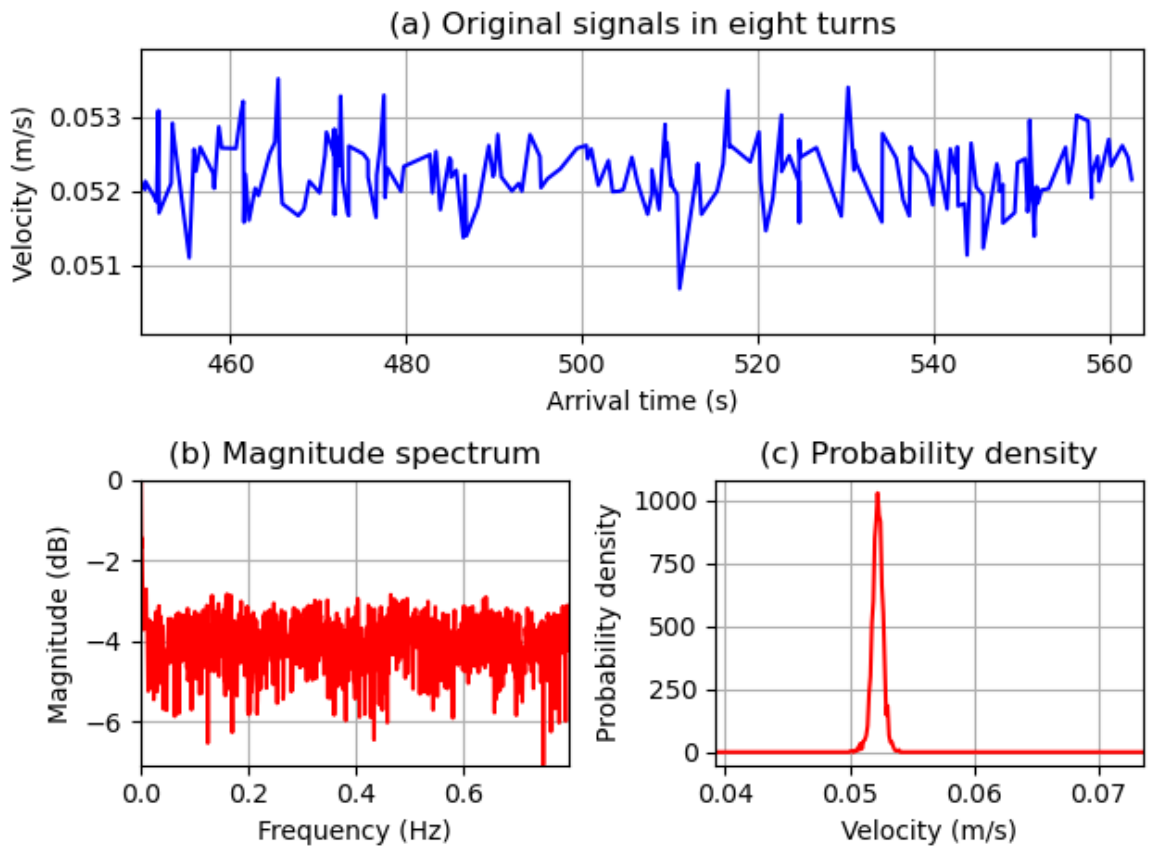

**Figure 5**. Signal segment at the measurement point $z_B = 247$ mm, $r = 55$ mm and its analysis

Fig. 5(a) shows the original signal fragments where the measured velocity oscillates within a very small range over time. Fig. 5(b) gives the magnitude spectrum almost like a white noise. Fig. 5(c) shows that the probability density is a good Gaussian distribution. Through the analyses of a large number of signals from different measuring points at different rotating speeds, we confirm that there are about 97.5% effective signals under 97% confidence level, and about 99.2% effective signals under 95% confidence level, the 95% confidence intervals (error bars) are lower than 0.1%.

Fig. 6 presents the velocity profile and the normalized angular speed at the axial center section $z_B = 247$ mm of the water body. The experimental results show that (i) the circumferential velocity increases monotonically with the radial coordinate, but not linearly; (ii) The normalized velocity profiles deviate from the rigid-body rotation distribution predicted by the strain rate model, maximal about 0.86% and independent of rotational speed, which are consistent with the results predicted by the slip model; (iii) As the angular speed increases, the relative mean square deviations at different measurement points slightly decrease and stabilize at around $\pm 2\%$.

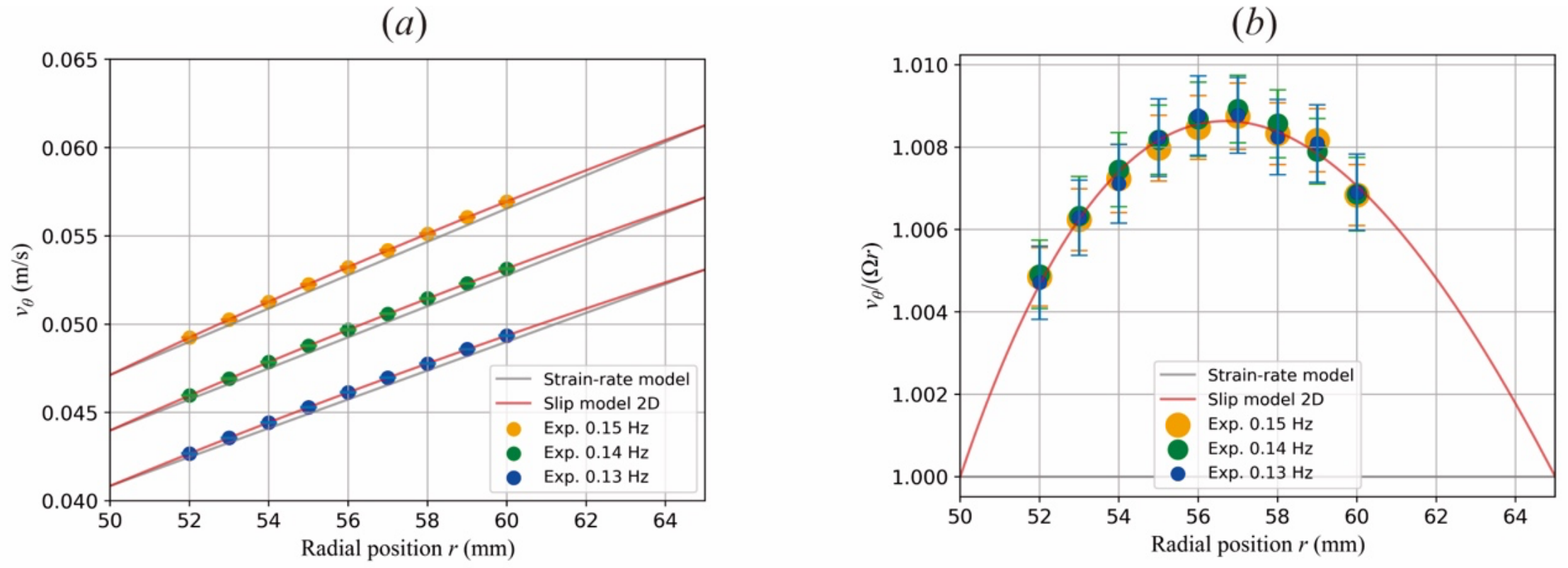

**Figure 6**. Experimental results at different angular speeds: (*a*) velocity profiles, (*b*) normalized velocity profiles.

Fig. 7 shows the placement of the laser head during the experiment. After extensive practice, we believe that the measurement scheme of laser incidence from the upper end can achieve better accuracy compared to that of laser incidence from the transparent outer cylinder wall.

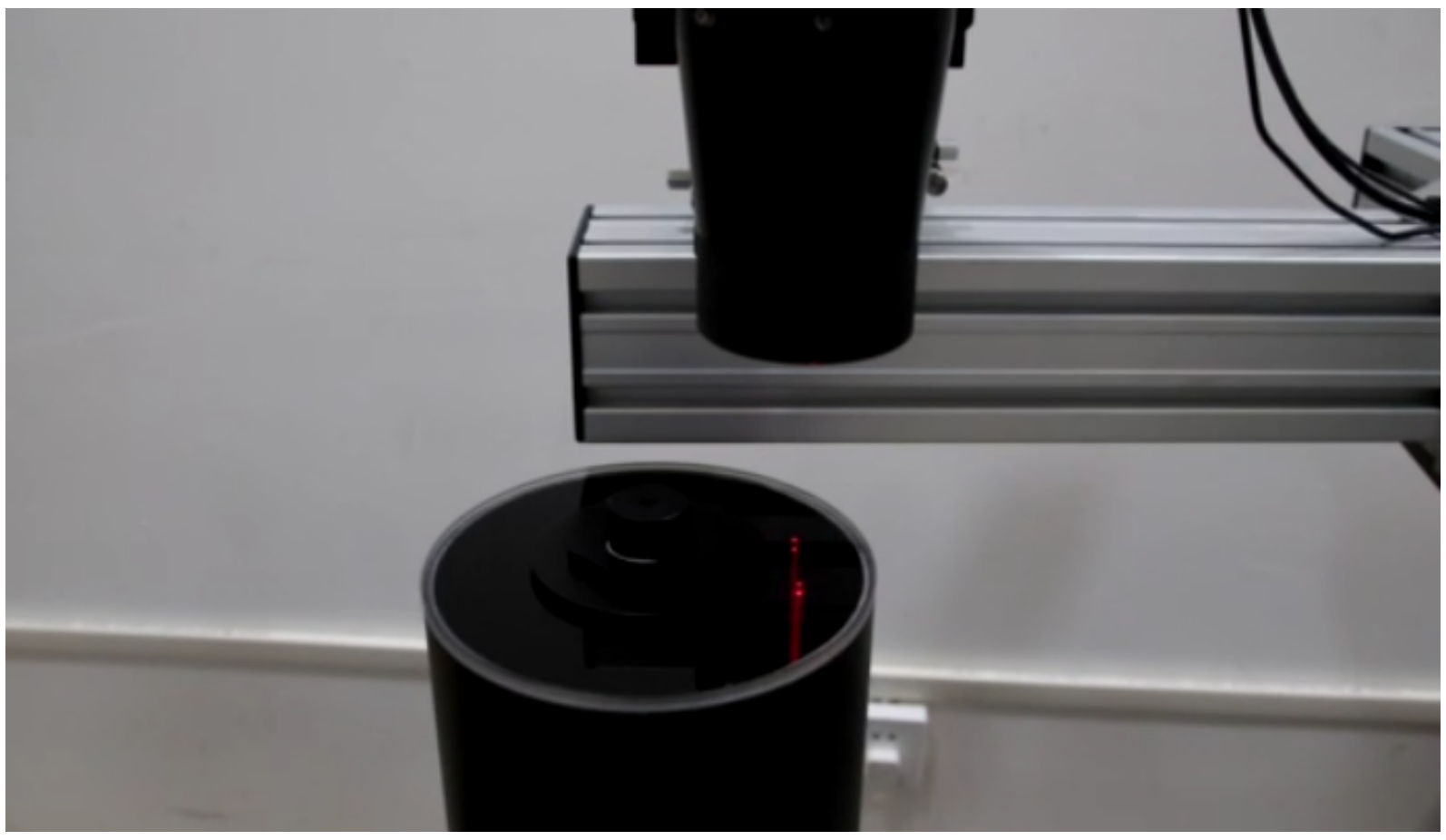

**Figure 7**. The placement of laser head during the experiment

## 3. SIMULATION

The predictions of two viscosity models in Fig. 6 do not consider the end effects. It is easy to prove that the rigid-body solution of the strain-rate model with ends, which is also verified in our simulation using the ANSYS Fluent software. But for the slip model, due to the constraint of ends, the circumferential velocity becomes the function of $r$ and $z$, so do the pressure. Then, the axial momentum equation yields that the axial velocity must not be zero, and further the continuity equation will confirm the appearance of the radial velocity. In summary, according to the slip viscosity model, the T-C laminar flow with ends must be three-dimensional (3D). we will demonstrate the 3D characteristics of the flow by simulation.

### 3.1. Numerical method and setting of geometric and flow parameters

In order to utilize the existing simulation software, we first introduce the expression

$$\nabla^2 V = \nabla^2 \sqrt{v_k v_k} = \nabla \cdot \left( \frac{1}{\sqrt{v_k v_k}} v_l \nabla v_l \right) = \frac{1}{V} v_k \nabla^2 v_k + \frac{1}{V} (\nabla v_k) \cdot (\nabla v_k) - \frac{1}{V^3} (v_k \nabla v_k) \cdot (v_l \nabla v_l) \quad (16)$$

to formulate the slip momentum equations (7) into a form such as the N-S equation plus an additional source, say

$$\partial_t v_i + v_k \partial_k v_i + \partial_i P - \nu(\nabla^2 v_i) = S_i, \tag{17}$$

where

$$S_i = \frac{\nu}{V^2} v_k (v_i \nabla^2 v_k - v_k \nabla^2 v_i) + \frac{\nu}{V^4} v_i v_l (v_l \nabla v_k - v_k \nabla v_l) \cdot \nabla v_k. \tag{18}$$

If denote by

$$a_1 = \frac{v_2 \nabla^2 v_3 - v_3 \nabla^2 v_2}{V^2}, a_2 = \frac{v_3 \nabla^2 v_1 - v_1 \nabla^2 v_3}{V^2}, a_3 = \frac{v_1 \nabla^2 v_2 - v_2 \nabla^2 v_1}{V^2}, \tag{19.1}$$

$$\boldsymbol{c}_1 = \frac{v_2 \nabla v_3 - v_3 \nabla v_2}{V^2}, \boldsymbol{c}_2 = \frac{v_3 \nabla v_1 - v_1 \nabla v_3}{V^2}, \boldsymbol{c}_3 = \frac{v_1 \nabla v_2 - v_2 \nabla v_1}{V^2}, \tag{19.2}$$

the source term $S_i$ can be simply written as

$$S_i = \nu \epsilon_{ijk} v_j a_k + \nu(\boldsymbol{c}_k \cdot \boldsymbol{c}_k) v_i. \tag{20}$$

The boundary conditions for two viscosity model are the same

$$v_\theta = \Omega R_1, r = R_1; v_\theta = \Omega R_2, r = R_2; z = 0, H, v_\theta = \Omega r. \tag{21}$$

The numerical example shown in Fig. 6- coming from the parameters

$$\Omega = 0.05 \text{ Hz}, 1 \text{ rad/s} \, (0.159 \text{ Hz}), 0.5 \text{ Hz}, R_1 = 50 \text{ mm}, R_2 = 65 \text{ mm}, H = 500 \text{ mm}. \tag{22}$$

This simulation is also carried out with the ANSYS Fluent solver attached by UDF (Zhu, 2019 [21]; Xu, 2020 [22]; Kong, 2025 [23]). Both two-dimensional axisymmetric swirl flow calculation and three-dimensional flow calculation are employed for cross-validation. The chooses of grid and algorithm are verified by solving the N-S equations underlying the strain-rate model. That is to say, the analytical solution of rigid-body rotation can be accurately reproduced through the numerical simulation. In order to guarantee the state of flow to be laminar, the angular speeds in the simulation are taken to be smaller values. In addition, we are able to utilize axial symmetry to reduce the computational domain by half.

### 3.2. Flow field display: end-wall effects or more?

In the following, we show the numerical results mainly using $\Omega = 1 \text{ rad/s}$ as the typical example.

For the confined T-C laminar flow, the numerical simulation based on the slip viscosity model reveals that the streamlines exhibit pronounced three-dimensional helical characteristics just like the forementioned analysis result from the equations, particularly near the ends, as illustrated in Fig. 8. In most domain away from the ends, the streamlines are almost circular. Near the lower end, the streamlines close to the outer cylinder wall are spiral downward, while those close to the inner cylinder wall are spiral upward. The streamlines located about 2.5 mm from the end are spiral horizontally from the outside inward.

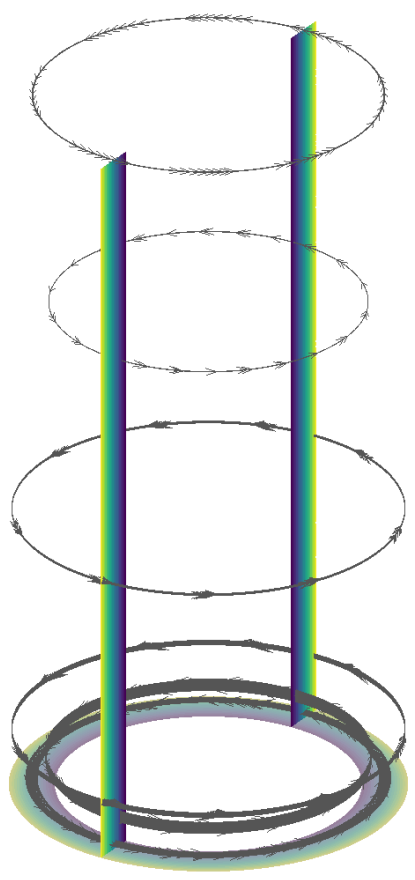

**Figure 8**. Three-dimensional helical streamlines

Although the flow is three-dimensional, the axial and radial motions are nearly three orders of magnitude smaller than circumferential motion, and are primarily concentrated near the ends. The axial motion is clearly divided into two regions by the radial position $r \approx 57.5$ mm an inner positive region and an outer negative region. The radial inward motion is concentrated in a thin layer near the ends, and then rapidly weakens towards the axial center symmetry plane, as shown in the Fig. 9.

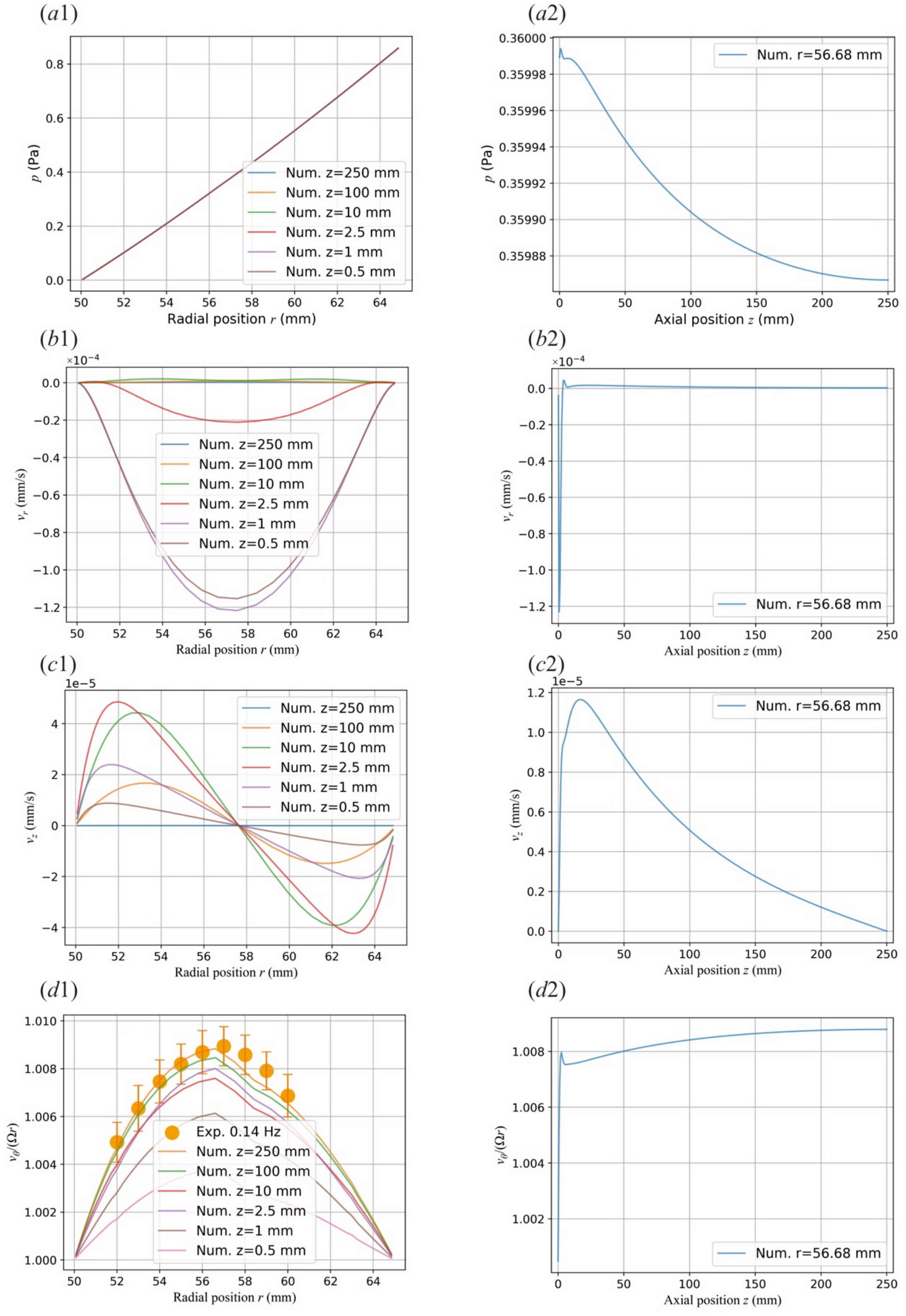

**Figure 9**. Pressure distribution and velocity profiles when $\Omega = 1$ rad/s: ($a$1)-($d$1) along radial direction at different axial positions, ($a$2)-($d$2) along axial direction at the radial position $r = 56.68$ mm.

The selection of axial positions $z = \{0.5, 1, 2.5, 10, 100, 250 \text{ mm}\}$ takes into account the changing characteristics, while the radial position $r = 56.68\ \mathrm{mm}$ is the point where the slip model predicts maximum circumferential velocity in the case without ends. Fig. 9($a1$, $a2$) show that the pressure change along the radial direction is almost independent of the axial position, while slight but drastic changes occur near the end. Fig. 9($b1$, $b2$) show that the relatively strong radial velocity appears a very thin region near the end, which is negative. Fig. 9($c1$, $c2$) show that the strong axial velocity also occurs a thin region near the end, but slightly thicker and positioned higher, facing the central symmetry plane in the inner side and facing the end in the outer side. Fig. 9($d1$, $d2$) show that the normalized circumferential velocity profile by the rigid-body rotation distribution exhibits a deficit outside its peak around the center section, and does not monotonically decrease to the boundary velocity towards the end, but rather exhibits a significant convex peak near the end.

In Fig. 10, we further present the normalized circumferential velocity under different angular speeds from simulation and experiment. For lower speeds, the stable experimental velocity profiles of center section coincide well with the numerical results, as shown in Fig. 10($a$); but when $\Omega = 0.5\ \mathrm{Hz}$ the numerical velocity profile becomes obvious lower. Fig. 10($b$) demonstrates the changes of circumferential velocity with the distance from the end. There are three typical features to be mentioned: (i) the velocity at the center section decreases with the speed; (ii) a velocity peak near the end appears due to the inward flow, which increases with the speed; (iii) the velocity profile exhibits a deficit outside its peak due to the outward flow, which is difficult to find the present experimental results.

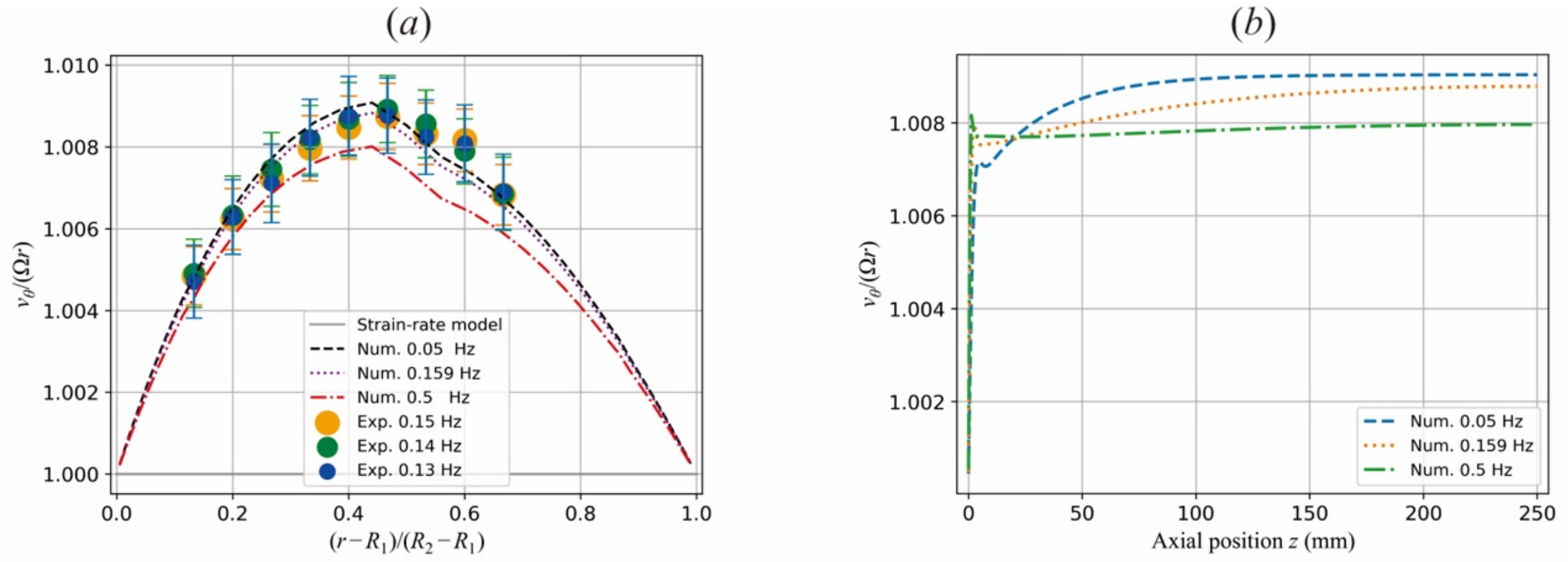

**Figure 10**. Comparison of numerical results with experimental results under different angular speeds: ($a$) normalized velocity profiles at the center section, ($b$) distribution of circumferential velocity along the axial direction at $r = 56.68\ \mathrm{mm}$ showing the effect of ends.

In order to obtain a better overall image of the flow field, we present the velocity component contours with angular speed $\Omega = 0.05\ \mathrm{Hz}$ in Fig. 11. For the circumferential velocity, the axial profile at $r = 56.68\ \mathrm{mm}$ and the radial profile at $z = 250\ \mathrm{mm}$ are further provided in Fig. 11($c$). In this case, the maximal circumferential velocity in the center section even exhibits a little larger than the maximal circumferential velocity without ends. The circumferential velocity profiles keep almost invariant until $z < 100\ \mathrm{mm}$, which is not so good when $\Omega = 0.05\ \mathrm{Hz}$. When the angular speed becomes larger, $\Omega = 0.5\ \mathrm{Hz}$, the peak value of circumferential velocity near the end may become the largest one (Fig. 10(b)).

From theoretical analyses and numerical simulations, it can be concluded that the T-C laminar flows of infinitely long cylinders are independent of Reynolds number $\mathrm{Re} = \Omega(R_2^2 - R_1^2)/2\nu$, and the difference between the two viscosity models increases with the increase of relative cylinder spacing $\eta = (R_2 - R_1)/R_1$. From the numerical simulations, we know that the velocity field of the T-C laminar flow with ends is invariant for the strain-rate viscosity model, but depends on the relative cylinder spacing, relative cylinder length $l = H/R_1$ and Reynolds number for the slip viscosity model. The preliminary experimental results under lower angular speeds or smaller Reynolds numbers indicate that the circumferential velocity profiles precisely measured by LDA certainly support the slip model than the strain-rate. In order to obtain the velocity profile closer to that of the T-C laminar flow without ends, we need a larger relative cylinder length and a lower Reynolds number (or a very low rotational speed when the relative cylinder spacing become larger), as well as precise geometry, strict control, and precise measurement. In the cases presented above, the Reynolds number should be lower than 1500.

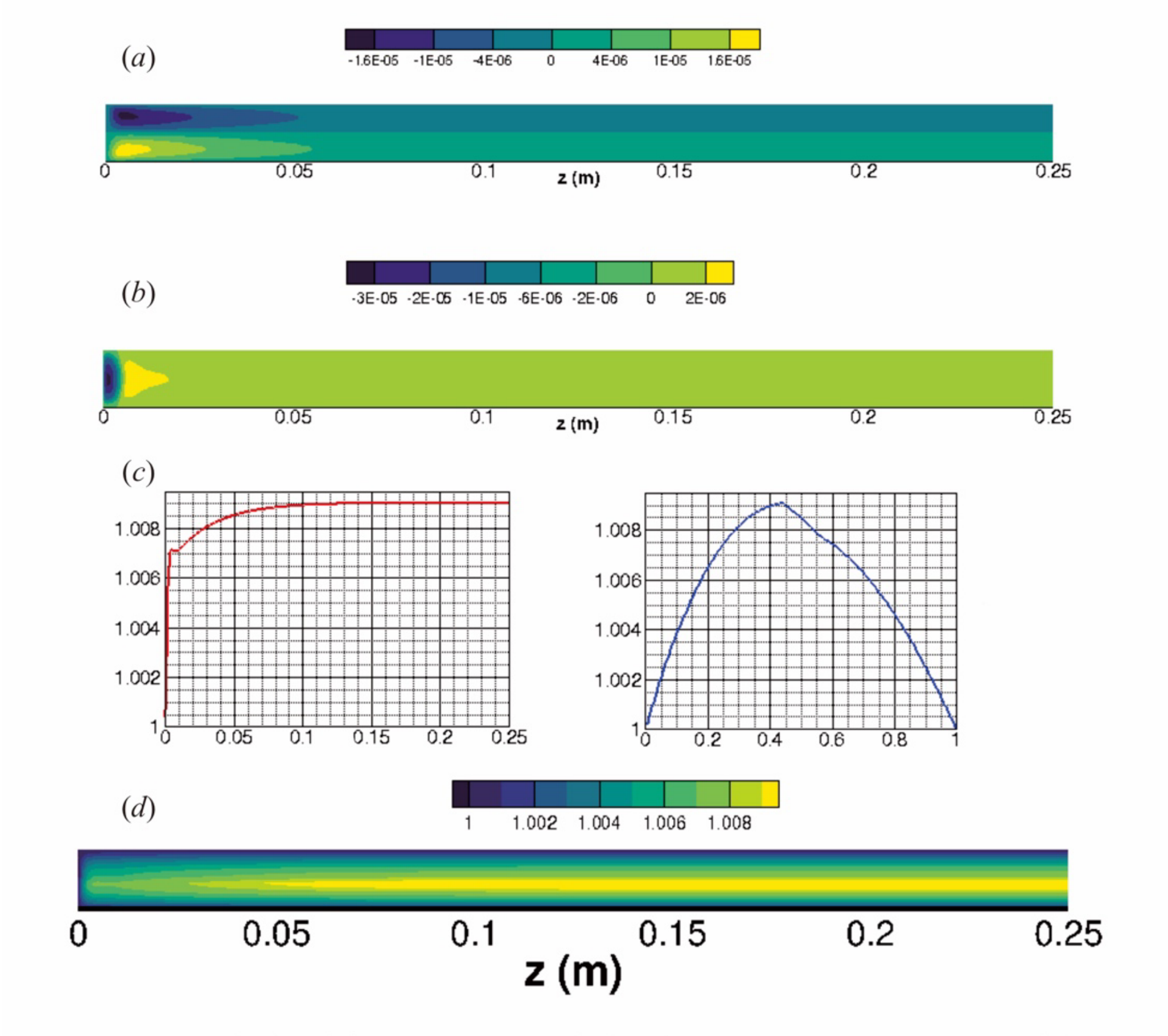

**Figure 11**. Contours of three velocity components and (*c*) typical profiles of normalized circumferential velocity: (*a*) axial velocity, (*b*) radial velocity, and (*d*) circumferential velocity.

## 4. SUMMARY AND OUTLOOK

In an experiment of T-C laminar flow experiment with the geometry $R_2 = 18\text{ in}, R_1 = 16\text{ in}, H = 54.7\text{ in}$, and the stationary inner cylinder while the outer cylinder and the ends rotate at different angular speeds ( $\text{Re} = \Omega R_2(R_2 - R_1)/\nu = 3000, 6000, 9000$ , respectively), Coles and van Atta (1966)[24] measured the circumferential velocity profile near the axial plane of symmetry with the platinum-rhodium hot wires and reported the obvious distortion from the predictions of the strain-rate model in the case without the ends. However, their formal object of the measurements was to establish a known flow for calibration of a hot-wire probe array for response to changes in pitch angle, yaw angle, and speed. But the fact was beyond their expectations, they unhesitatingly explained this deviation as the influence of the end-walls, without suspecting that the predictions of the conventional viscosity model were flawed.

In the study of this paper, we chose a simpler T-C flow configuration, where the prediction of the strain-rate viscosity model can be unaffected by the presence of the ends. The results of experimental measurements, namely the circumferential velocity profiles of the center section, deviate from the rigid-body rotation distribution, but meet well with the predictions of the slip viscosity model. The numerical simulations have revealed the influence mode and scope of the end-wall effects, which can be used to guide the experiments in the future. The weak three-dimensionality of flow greatly changes the circumferential velocity profiles along the axial direction, where the Reynolds number plays a significant role.

Once the thinking framework of the strain-rate model is broken, there is a lot of work to be done. The prediction deviations of classical flow theory in curved laminar flows will be repeatedly verified through experiments. Just like from the geocentric theory to the heliocentric theory, the transion from the strain-rate viscosity model to the slip viscosity model will also bring a jump of understanding on real flow and promote the solution of turbulence problems.

**Acknowledgments:** The authors thank to all the people who have provided help and encouragement in the process of this research. W.N.Z. acknowledges the support of the Department of Engineering Mechanics, Nanchang University, especially the support of Prof. Chun Zhang, the dean of the department.

**Author Declaration:** The authors report no conflict of interest.

**Data availability:** The data that support the findings of this study are available from the corresponding authors on reasonable request.

**Author Contributions:** W.N.Z. proposed the slip viscosity model, put forward the idea of falsification experiment and supervise the simulation. W.N.Z. and X.H.Z. conceived the experiment, and together with X.S.C. determined the preliminary experimental scheme. J.W., under the supervision of X.S.C. in University of Shanghai for Science and Technology, constructed the experimental platform of the T-C flow, but failed to carry out the accurate measurement of velocity profile using the PIV technology. Based on the experimental platform built up by J.W., J.H., under the supervision of P.L.Z. in Nanchang University, carried out the accurate measurement of velocity profile using the LDA technology, while Q.C.K. participated in later experiments. W.N.Z. conducted the whole experimental processes as the Master supervisor of J.W. (2018-2020) and J.H. (2021-), including the organization of many discussions with X.S.C. and X.H.Z. during 2018-2020, with P.L.Z., J.W. and X.H.Z. during 2021-2022, proposing the modification and analysis frame of experimental data, and writing the manuscript with contributions from the rest of the authors. J.W. made equally important contribution in the experiment as J.H.